# Engineering a full gamut of structural colors in all-dielectric mesoporous network metamaterials

Alejandra Ruiz-Clavijo[1], Yoichiro Tsurimaki[2], Olga Caballero-Calero[1], George Ni[2], Gang Chen[2], Svetlana V. Boriskina[2*], Marisol Martín-González[1,2*]

[1] IMN-Instituto de Micro y Nanotecnología (CNM-CSIC), Isaac Newton 8, PTM, E-28760 Tres Cantos, Madrid, Spain.
[2] Department of Mechanical Engineering, Massachusetts Institute of Technology, Cambridge, MA 02139, USA

**Abstract**
Structural colors are a result of the scattering of certain frequencies of the incident light on micro- or nano-scale features in a material. This is a quite different phenomenon to that of colors produced by absorption of different frequencies of the visible spectrum by pigments or dyes, which is the most common way of coloring used in our daily life. However, structural colors are more robust and can be engineered to span most of the visible spectrum without changing the base material, only its internal structure. They are abundant in nature, with examples as colorful as beetles covers and butterfly wings, but there are few ways of preparing them for large-scale commercial applications for real-world uses. In this work, we present a technique to create a full gamut of structural colors based on a low-cost, robust and scalable fabrication of periodic network structures in porous alumina as well as the strategy to theoretically predict and engineer different colors on demand. We experimentally demonstrate mesoporous network metamaterial structures with engineered colors spanning the whole optical spectrum and discuss their applications in sensing, environmental monitoring, biomimetic tissues engineering, etc.

Colors are abundant in nature, and color perception is an invaluable tool for animals and humans alike that aids in communication, navigation, and survival strategies. Evolution has played a major role in the development of different colors in animals, plants, etc., and the human eye is designed to distinguish them. The perceived color of any material is a combination of its selective absorption and reflection of different wavelengths of incident light, and of specific wavelength spectra of light originating from different light sources, such as the sun or a lightbulb. Colors can typically be classified as either pigmentary or structural by their mechanism of formation on a material level. Pigmentary colors are based on frequency-selective light absorption in different materials. Materials transparent to the visible light in their bulk form can also exhibit color if they are micro- or nano-structured to resonantly scatter light of a specific wavelength. Such frequency-selective light reflection due to internal optical resonances in the structured material may result in its visible coloring, known as structural color[1,2]. Since the perceived color of structured materials does not directly depend on their absorptance, it opens up enormous opportunities for color engineering with transparent and lossless materials as well as for combining multiple functionalities within the same material.

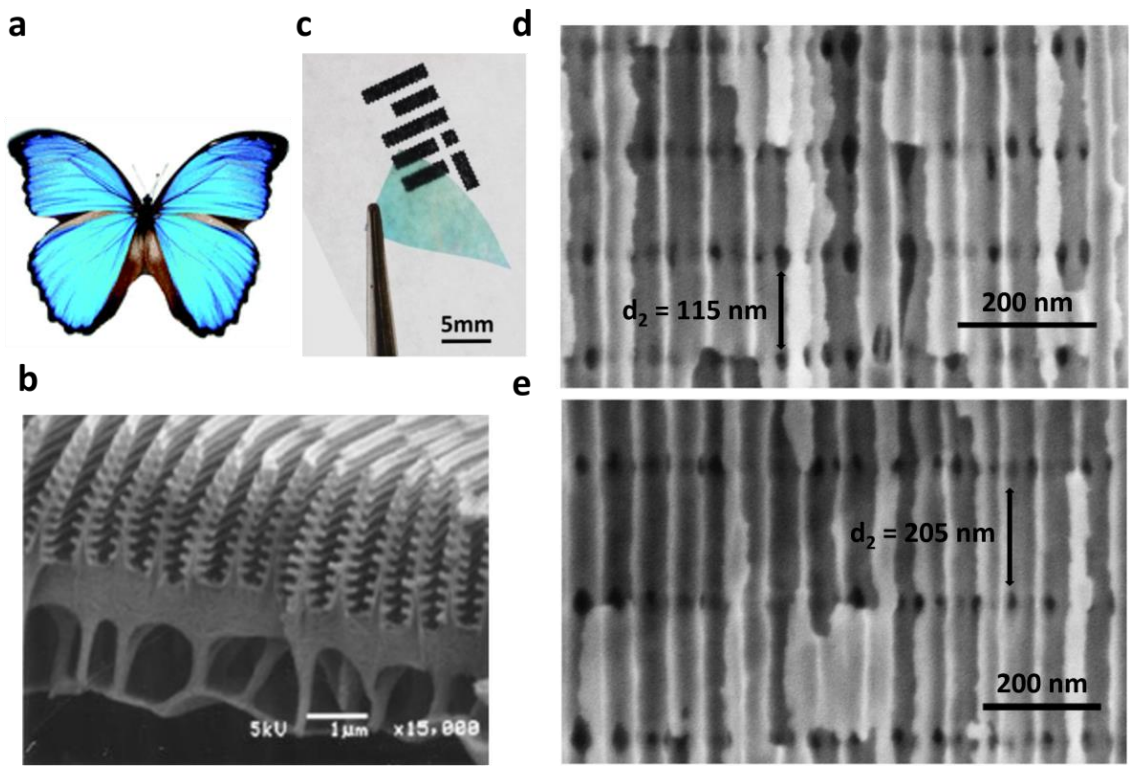

**Figure 1. Natural and artificial 3D mesoporous network metamaterials enable structural color formation.** (a) The blue color of the *Morpho* butterfly is formed by spectrally-selective light reflection from the mesoporous structure of its wings (reproduced with permission from[3]). (b) SEM image of the *Morpho* wing internal composition (reproduced with permission from Shinya Yoshioka, Osaka University[4]). (c) Blue color formation on an artificial 3D mesoporous network metamaterial sample. (d,e) SEM images of two of the fabricated samples, which exhibit a regular pattern of horizontal layers of varying thickness with vertical pores separated by narrow fixed-thickness layers with intersecting longitudinal and transverse pores.

There are many examples of structural color formation in nature, both in the inorganic world and in living organisms including birds, plants, insects, etc.[1,5–9]. One of the most studied is the microstructure present in butterfly wings, which is based on a periodic arrangement of lamellae, ribs, and ridges (Fig. 1a,b)[6,10]. It provides important survival features, such as hydrophobicity and aerodynamics, in addition to the typical iridescence colors that originate from the interaction of light with the mesoporous scales. Structurally-colored structures have been extensively studied for practical applications, such as biochemical sensors and photodetectors with colorimetric transduction mechanism[10–15]. To observe structural color in artificial materials, they need to be engineered to exhibit internal micro- and nano-scale order. This can be achieved by either using natural templates such as butterfly wings and modifying them by coating, filling, or oxidation[10–12], or via top-down or bottom-up nano-fabrication.

Top-down techniques can be used to produce three-dimensional (3D) structures by stacking two-dimensional (2D) layers on top of each other, although this process is often expensive and has low throughput[16]. Layered structures yield structural color in fibers[17–19], but fiber fabrication processes are not easily transferable to the planar materials. Holographic lithography can produce ordered 3D structures with the features in the range of hundreds of nanometers[20], yet requires a rather complicated setup and is currently limited to laboratory scale fabrication. Other novel lithography-based techniques, such as high precision 3D laser cutting (e.g. Nanoscribe GmbH), are also limited to relatively small areas, below 4 in[2], and have a high cost, in the range of hundreds of euros for 100 $\mu m^2$ pattern. Focused Ion Beam (FIB) techniques offer nanometer-scale resolution[21], yet suffer from high cost (over a hundred euros per hour of FIB utilization, plus materials costs), require several steps to process a 3D structure, and are limited to areas on the order of 100 $\mu m^2$[22]. Structurally-colored samples that are microscopic in size require complicated optics to observe colors, which limits their practical use as naked-eye sensors and detectors. Finally, structurally-colored planar multi-layer materials only interact with the environment through their flat external surfaces, which severely limits their color change sensitivity.

The bottom-up fabrication processes are typically less expensive than the top-down ones, and allow larger-scale fabrication, although they often rely on the use of high-cost materials such as noble-metal nanoparticles[23–28], and typically lack internal periodicity over large surface areas or large thicknesses. For example, mesoporous metamaterials grown from dealloying can only exhibit color in combination with additional dielectric coatings[29]. Nevertheless, several bottom-up techniques have been developed that give rise to structural coloring arising from the material internal structure. They still suffer from various drawbacks, such as difficulties with large-scale processing and formation of free-standing films associated with artificial opals fabrication[30–32], necessity to combine different growth techniques, including colloidal assembly, electrochemistry and atomic layer deposition[33], as well as issues with stability, shrinkage, and delamination associated with the use of self-assembling block copolymers[34,35]. Furthermore, the colors generated via bottom-up techniques, especially those that depend on the individual particle resonances, may not exhibit a large palette, which limits their applications.

As such, there is still a lack of simple, low-cost, high-throughput and scalable methods for fabrication of robust structurally-colored metamaterials, which would simultaneously provide nano- and micro-scale structures for whole-gamut color formation, large surface-to-volume ratios for improved sensing performance, and large areas for real-world applications. Currently, there are methods to either produce structural coloring reproducing almost the whole palette, but at a high cost and with complicated fabrication techniques,[36–38] or with simpler techniques that fail to cover the whole gamut of visible colors[29,39–41]. To address these limitations, we present design and fabrication strategies to develop large area alumina mesoporous structures with three-dimensional (3D) internal structure via anodization of aluminum to controllably create sub-wavelength periodic features in nano-porous dielectric materials (see Fig. 1d,e). This method is well suited for low-cost (in the laboratory, less than 0.3 euros per mm[2]), large-scale (m[2]) fabrication and allows engineering of a whole gamut of visible colors in a single lossless dielectric material.

The colorimetric optical response of the fabricated structures is formed as a result of destructive and constructive interference of light reflected from multiple interfaces within the metamaterial. The periodic nature of the metamaterial structures makes them tunable to cover the whole gamut of visible colors. Anodized porous alumina materials have already been explored for sensing and structural coloring applications, providing insight into their color formation mechanisms[42–47]. However, in most cases, the color was formed by engineering interference in thin films of porous alumina with various substrates or overcoats[42–44,46], which significantly limited the range of accessible hues[42]. Even AAO structures with two-dimensional internal periodic order typically exhibited only one reflection band, tunable through a part or the whole visible spectrum via variation of process parameters[45,48]. The colorimetric response of 3D porous anodized alumina structures studied in this manuscript can be formed by either one or multiple reflection bands. This allows for tuning of their structural color within the full range accessible by standard Red, Green, Blue (RGB) display technologies, the creation of metamer color filters for security applications, and high sensitivity of their colorimetric response to environmental changes. The nanoporous structure of our 3D samples enables fast and easy penetration of liquids and gases throughout the sample volume and the corresponding change in their perceived color. We demonstrate structural color sensitivity to environmental changes (i.e., material wetting), which can be easily detected by the naked eye.

**Results and Discussion**

We prepared anodic alumina mesoporous meta-material samples, which exhibit longitudinal pores of around 40 – 50 nm in diameter extending from the top down perpendicular to the aluminum surface, and transversal pores of around 20 – 25 nm in diameter, which connect longitudinal pores (see Fig. 1d,e and Supplementary Fig. S1). The horizontal layers with transversal pores can be formed with pre-defined periodicity in the vertical direction. The fabrication method – which we previously developed to fabricate nanoporous materials for thermoelectric applications[49,50] – is schematically shown in supplementary Fig. S2. It is based on sequentially pulsed anodization and chemical etching steps and enables complete control over the distances between the periodically formed horizontal layers with intersecting longitudinal and transversal pores (see Methods). Typical scanning electron microscopy (SEM) images of fabricated samples are shown in Figs. 1d,e, and reveal their intricate internal mesoscale structure with the periodicity on the length scale comparable to that present in the butterfly wings (compare to Fig. 1b). This periodicity varies from sample to sample as desired and defines the sample colorimetric response. Samples with periodically repeating transversal pores are effectively one-dimensional photonic crystals[51,52] with internal nanoporous structure and can be engineered to exhibit spectrally-selective coherent scattering of light of select colors.

Figure 2a shows a schematic of the model periodic structure we used to predict and tune the optical response of mesoporous dielectric meta-materials. Each horizontal layer is characterized by an average porosity value. High-porosity layers are the layers with intersecting longitudinal and transverse pores, while low-porosity layers are the layers with only longitudinal pores in the fabricated samples (see Figs. 1d,e, and Supplementary Fig. S1). Bulk alumina is a dielectric material transparent across the visible spectral range (Supplementary Fig. S3). In turn, nanoporous alumina samples can be characterized by effective refractive indices of high- and low-porosity layers, which in this work were calculated via the Maxwell-Garnett theory (see Methods and Supplementary Fig. S4). The reflectance spectra of periodic structures have been obtained by using the robust semi-analytical transfer matrix method (see Methods and Supplementary Fig. S5). To convert the reflectance spectra into the perceived color prediction for each structure, we used the CIE color space chromaticity diagram, introduced in 1931 by the International Commission on Illumination (CIE) (see Supplementary Fig. S6)[53]. Human eyes have three types of light-sensitive cone cell receptors, which have different sensitivities to different wavelengths in the solar spectrum (Fig. 2b). Combining these sensitivities with the reflectance spectra of different samples, we can predict the perceived color by characterizing it with two coordinates within the CIE color space (see Methods).

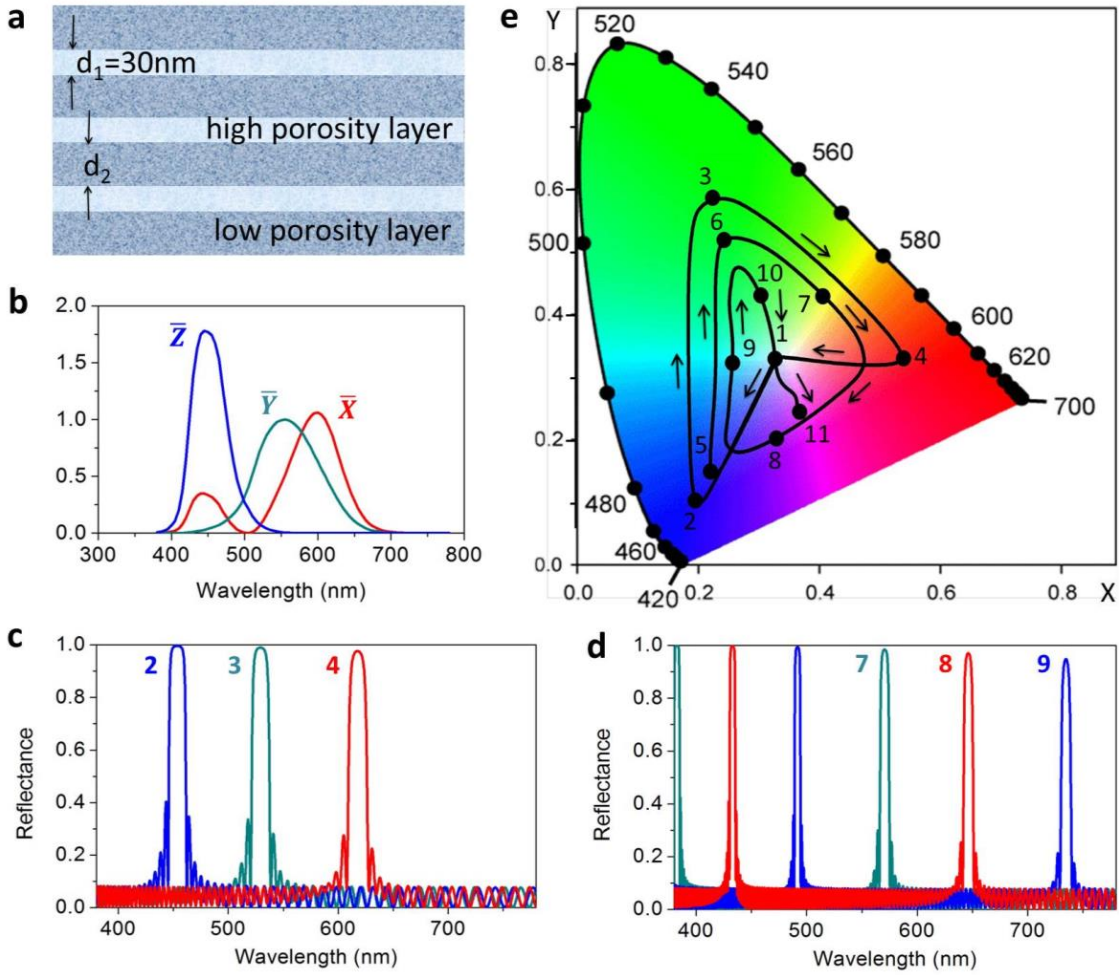

**Figure 2. Engineering structural color in 3D porous network metamaterials.** (a) Schematic of the periodic metamaterial with periodic composition of high-porosity and low-porosity layers characterized by the corresponding effective index values. (b) CIE standard observer color matching functions. (c,d) Calculated reflectance spectra of periodic (150-layers thick) structures shown in panel (a) with varying thickness of the low-porosity layer (with only longitudinal pores), $d_2$, and fixed thickness of the high-porosity layer (with both longitudinal and transverse pores), $d_1 = 30$nm. The structural parameters of the samples labeled with numbers in (c-e) are summarized in Table 1. (e) CIE color space representation (background) with the colors theoretically predicted for structures in Table 1 shown as dots. The black curve shows the color predictions for structures with continuously increasing low-porosity layer thickness, and the arrows illustrate the predicted color evolution direction.

Reflectance spectra of periodic structures in Fig. 2a predicted via modeling are shown in Figs. 2c and d for several gradually increasing values of sample periodicity, summarized in Table 1. The thickness of the high-porosity layer was fixed at 30nm for all the samples (i.e., $d_1 = 30$nm), which corresponds to the experimentally-observed situation (see Figs. 1d,e). In turn, the porosities of the two layers used in calculations were chosen as 65% (for the low-porosity layers with longitudinal pores only) and 80% (for the high-porosity layers with both longitudinal and transverse pores), respectively. These values were found by training the model on measured spectra of multiple fabricated samples to find the unique porosity values that match all the experimentally-observed spectral bands of all the samples in the training set (see Methods). The high-reflectivity peaks in the reflectance spectra in Figs. 2c,d correspond to the formation of the photonic band gaps in the material, where light propagation through the material is forbidden, and strong reflection occurs instead[51,52]. It can be seen that with the increased period of the structure (i.e., with increased distance $d_1 + d_2$), the first-order reflection band red-shifts (Fig. 2c), and the higher-order bands eventually move into the visible frequency range (Fig. 2d). Table 1 lists the spectral positions of all the reflectance peaks of structures labeled in Fig. 2e in the visible and near-infrared range, although some of the peaks (in the infrared and ultraviolet range) do not contribute to colors perceived by the human eye.

**Table 1.** Structural parameters of the periodic mesoporous structures with varying color predicted by modeling and shown in Fig. 2. The reported peak positions are rounded to the nearest whole wavelength value measured in nm.

| Sample # | 1 | 2 | 3 | 4 | 5 | 6 | 7 | 8 | 9 | 10 | 11 |
|---|---|---|---|---|---|---|---|---|---|---|---|
| $d_1 + d_2$, nm | 130 | 180 | 210 | 245 | 360 | 415 | 455 | 510 | 580 | 650 | 730 |
| $d_2$, nm | 100 | 150 | 180 | 215 | 330 | 385 | 425 | 480 | 550 | 620 | 700 |
| Color | white | blue | green | red | blue | green | yellow | purple | cyan | green | pink |
| Peaks, nm | | | | | | | | | | | |
| 5th order | | | | | | | | | | 333 | 373 |
| 4th order | | | | | | | | | 371 | 415 | 465 |
| 3rd order | | | | | | 353 | 383 | 433 | 492 | 551 | 618 |
| 2nd order | | | | | 457 | 527 | 571 | 646 | 735 | 824 | 925 |
| 1st order | 328 | 454 | 529 | 618 | 910 | 1049 | 1150 | 1290 | 1467 | 1643 | 1845 |

The coordinates on the CIE colorimetric scale corresponding to the calculated reflection spectra of the structures with varying periodicity listed in Table 1 are shown as black dots in Fig. 2e. They are connected by a black line, which illustrates the evolution of the predicted color with the gradual change of the structure period. Our modeling predicts that by simply varying the period of the structure, we can engineer samples exhibiting structural colors spanning the whole gamut of visible colors achievable with standard Red/Green/Blue (RGB) technologies (see Supplementary Fig. S6). Structures with the smallest periodicities $d_1 + d_2 \leq 130$nm) are expected to exhibit no color due to the absence of the forbidden gaps in the visible frequency range (we label them as white, while they are actually nearly completely transparent just like the bulk alumina). With the increase of the structure period, the forbidden band red-shifts (see Table 1), and blue coloring of structures is predicted to emerge, which evolves into green and then red as the period is increased and the reflection band moves through the visible range.

As the period of the modeled structures is increased even further, the predicted color converges gradually towards white in the center of the color scale. This is a result of additive color formation due to reflection within multiple forbidden frequency bands, and it is in stark contrast to the pigment color formation mechanism. Pigments produce black color if mixed, owing to additive absorptance of mixed single-color substances[54]. In turn, mixing of reflected light of different wavelengths spanning the whole visible spectrum yields white color (Supplementary Fig. S7).

Based on the model predictions for structural color formation in periodic mesoporous network metamaterials, we fabricated a set of samples that reflect light within a given wavelength range, with the goal to engineer the whole gamut of colors in the visible spectrum. The structural parameters of the fabricated set of samples are summarized in Table 2. Table 2 shows both the target values of parameters used in the design of the structures, and the actual values measured once the samples were fabricated. The actual thicknesses of the high- and low-porosity layers reported in Table 2 were extracted from the high-resolution SEM images of the fabricated samples (see Supplementary Fig. S8). Then, the reflectance of the fabricated samples with different transverse nanopore periodicity was measured at near-normal angle (82° to the surface plane). All the samples exhibited multi-band reflectance spectra, which shifted to longer wavelengths as the transverse pore periodicity of the sample was increased (see Figs. 3a-c, Supplementary Fig S9 and Table 2), in perfect agreement with the model predictions. The color coordinates calculated based on the experimental spectra were overlapped with the CIE color diagram in Fig. 3d. These colors can be clearly seen by the naked eye, and are shown in optical photographs of samples in Figs. 3e-g, which were taken at an angle normal to the sample plane.

**Table 2.** Structural parameters and colors of the fabricated samples. (*: This value corresponds to the longest-wavelength reflectance peak that appears in the measured range)

| Sample # | AAO3D1 | AAO3D2 | AAO3D3 | AAO3D4 | AAO3D5 |
|---|---|---|---|---|---|
| Designed periodicity ($d_1 + d_2$, nm) | 495 | 655 | 175 | 145 | 690 |
| Experimental periodicity ($d_1 + d_2$, nm) | 497 | 654 | 173 | 150 | 689 |
| Designed low-porosity layer thickness ($d_2$, nm) | 465 | 625 | 145 | 115 | 660 |
| Actual low-porosity layer thickness ($d_2$, nm) | 465 | 622 | 144 | 115 | 658 |
| Fitted porosity values (%) | 67/80 | 67/80 | 64/80 | 63/80 | 65/80 |
| Measured peak positions (nm) | | | | | 352 |
| 5th order | | 331 | | | 436 |
| 4th order | | 413 | | | 580* |
| 3rd order | 403&423 | 547* | | | |
| 2nd order | 603&634* | 820 | | | |
| 1st order | | | 439* | 386* | |
| Color (normal angle, in air) | green | red | blue | white | pink |
| Fitted peak positions (nm) | | | | | |
| 5th order | | 331.4 | | | 350.8 |
| 4th order | | 412.2 | | | 436.5 |
| 3rd order | 414.6 | 548.1 | | | 580.0 |
| 2nd order | 619.2 | 819.0 | | | 867.6 |
| 1st order | 1234.6 | 1634.6 | 439.2 | 381.5 | 1730.6 |

To design samples AAO3D1-AAO3D5, we used the same average porosity values as for the model structures in Table 1, i.e., 65% porosity of the layers with only longitudinal pores, and 80% porosity of the layers with intersecting pores. All the fabricated samples exhibited layer thicknesses within a few nanometers of the target thicknesses (see Table 2). To account for fabrication imperfections, including slight variations in porosity and layer thicknesses across the sample, some of the measured spectra were best fitted with slightly different values of porosity (see Table 2). Supplementary Fig. S10 illustrates the process of the spectral design and post-fabrication fitting for sample AAO3D2, and reveals a very good overlap between the model, experimental, and fitted spectra of this structure. The observed broadening of the experimental peaks is a result of variation in the layer thicknesses, which becomes more pronounced in thicker samples, and affects higher-order bands stronger than the first-order ones (Supplementary Fig. S11). Most importantly, the resulting observable colors of all the samples matched the model predictions using average porosity values of 65/80 (see e.g. Fig. 2d), confirming the predictive power of the developed model. For comparison, we also fabricated porous alumina samples without periodic transverse pores, i.e. with only longitudinal pores spanning throughout the sample, and observed no peaks in their reflectance spectra (Supplementary Fig. S12). The reflectance spectra of these samples were fitted well by simulations assuming 65% uniform porosity value (Supplementary Fig. S12). This clearly illustrates that the multi-band reflectance spectra in periodic samples originate from the formation of the photonic forbidden bands.

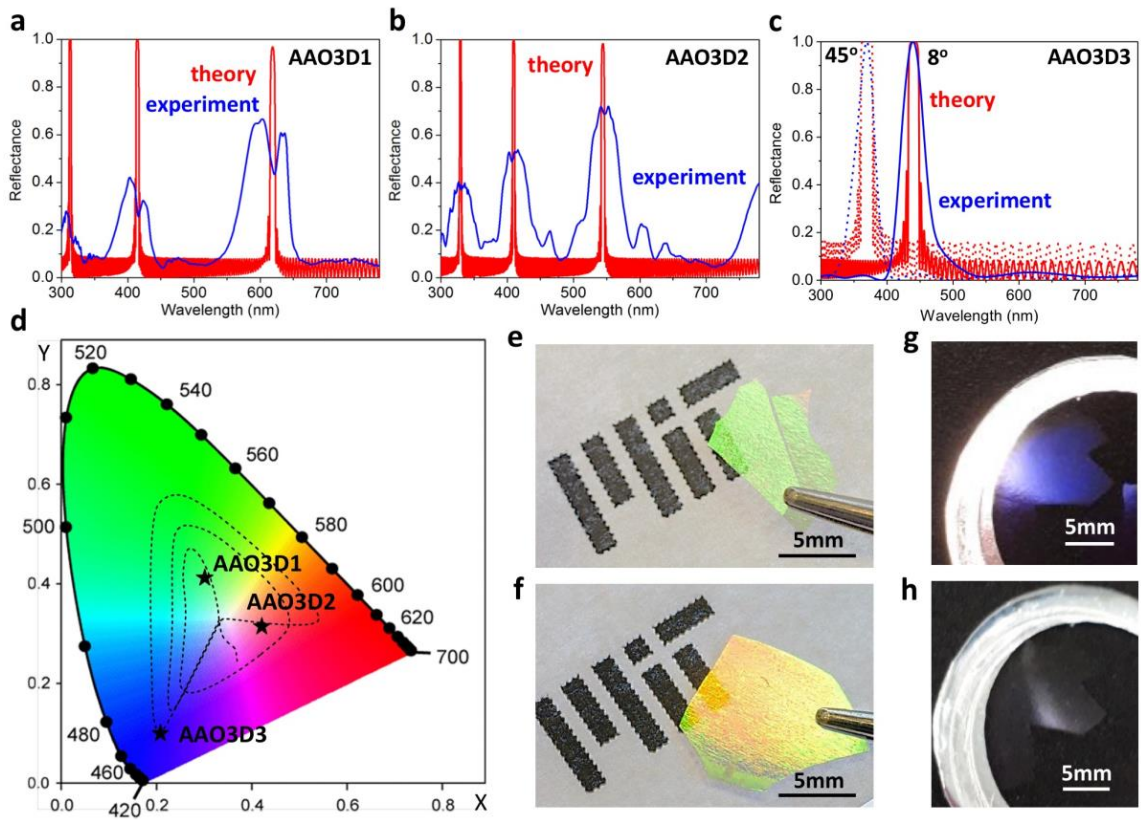

**Figure 3. Experimental demonstration of the color formation in porous 3D network metamaterials.** (a-c) Experimental (blue solid lines) and numerically-fitted (red solid lines) reflectance spectra under near-normal illumination of several samples with varying periodicity optimized to exhibit different colors. The corresponding experimental and theoretical spectra under oblique illumination at 45º to normal are shown in (c) as the blue and red dash lines, respectively. (d) Colors of samples AAO3D1-AAO3D3 on the CIE color diagram. (e-g) Photographs of the fabricated samples AAO3D1 (e), AAO3D2 (f) and AAO3D3 (g) at the normal viewing angle revealing the structural colors formation. (h) Photograph of sample AAO3D3 from the oblique viewing angle of 60 degrees from the normal to the surface exhibiting no color.

Unlike the colors of samples AAO3D1 and AAO3D2, the blue color exhibited by sample AAO3D3 is only visible at a normal viewing angle and disappears if viewed at oblique angles above 40 degrees with respect to the normal. This is due to forbidden gaps shifting as a function of the illumination angle (i.e., photon momentum), which causes the perceived color to change. Sample AAO3D3 becomes transparent (Fig. 3h and supplementary Fig. S13) if viewed at an angle because of its reflectance band blue-shifts and moves out of the visible spectrum range. The measured and calculated blue-shifted spectra of sample AAO3D3 for a viewing angle of 45º are shown as the dotted lines in Fig. 3c. Supplementary Fig. S13 illustrates the evolution of the reflectance spectrum and the resulting colorimetric signature of sample AAO3D3 as a function of the observation angle.

The perceived color of fabricated porous network meta-material samples is sensitive not only to changes in their periodicity or the observation angle, but also to changes in their environment. Color sensitivity to environmental changes can be exploited for engineering sensors with colorimetric transduction. Depending on the sample composition, the colorimetric readout can be in the form of (i) color appearing or disappearing (i.e., material switching between being either transparent or colorful), or (ii) distinct color change. We qualitatively demonstrate these effects in Fig. 4, where color sensitivity to the meta-material wetting is illustrated. Due to the high porosity of the meta-material, its immersion in water (or any other liquid or gas) will change light propagation through every layer in the bulk, and not just at the interface between the top layer and the environment. This change in the effective refractive index results in dramatic spectral shifts of the reflection bands, which in turn changes the color signature. Theory predicts that more striking colorimetric effects can be expected in structures where environmental changes result in a reflectance band either moving into or out of the visible range. We used this understanding, together with the results of the systematic design of the colorimetric response of samples with varying periodicity shown in Fig. 2e, to fabricate meta-material samples promising robust colorimetric readout (samples AAO3D4 and AAO3D5 in Table 2).

Figure 4a compares the predicted and measured reflectance spectra of sample AAO3D4. The sample is transparent to the eye (Fig. 4b) because its only reflectance band is well in the ultraviolet spectral region. It is also transparent for all incident and collection angles (supplementary Fig. S14). When immersed in water, however, the sample acquired blue color (Fig. 4c) owing to the red-shift of the reflectance band into the blue part of the visible spectrum (as predicted by theory and shown in Fig. 4a, red dotted line, when the air in the pores is replaced by water). Sample AAO3D5 has a much larger inter-pore distance than sample AAO3D4, and its spectrum has two reflectance peaks in the visible spectrum (Fig. 4d), which combine to yield red color (Fig. 4e), and one peak slightly blue-shifted from the visible wavelength range. Upon wetting with water, the sample color changed from red to green (Fig. 4f) owing to the significant spectral shift of all the three reflectance bands (Fig. 4d, red dotted line). Figures 4g and 4h illustrate the change of the color coordinates of samples AAO3D4 (g) and AAO3D5 (h) on the CIE color diagram, which occurs when the metamaterial is wetted.

Finally, since the base network material (i.e., alumina) is transparent across the whole visible range, the mesoporous samples can exhibit significant color variations if placed on different substrates. The part of the spectrum reflected by the substrate will propagate back through the sample and will mix with the reflection from the meta-material to produce a different color. Figure 4i illustrates the change of perceived color of sample AAO3D5, which was placed on top of a 100 US dollar bill. The portion of the sample AAO3D5 that was on top of a purple stripe prominently changed its perceived color from red to green. This feature makes our structures potentially useful for security, anti-counterfeiting, camouflage detection, and food safety applications[55–57]. They can also help to discriminate metamer materials exhibiting colors indistinguishable by a naked eye[58].

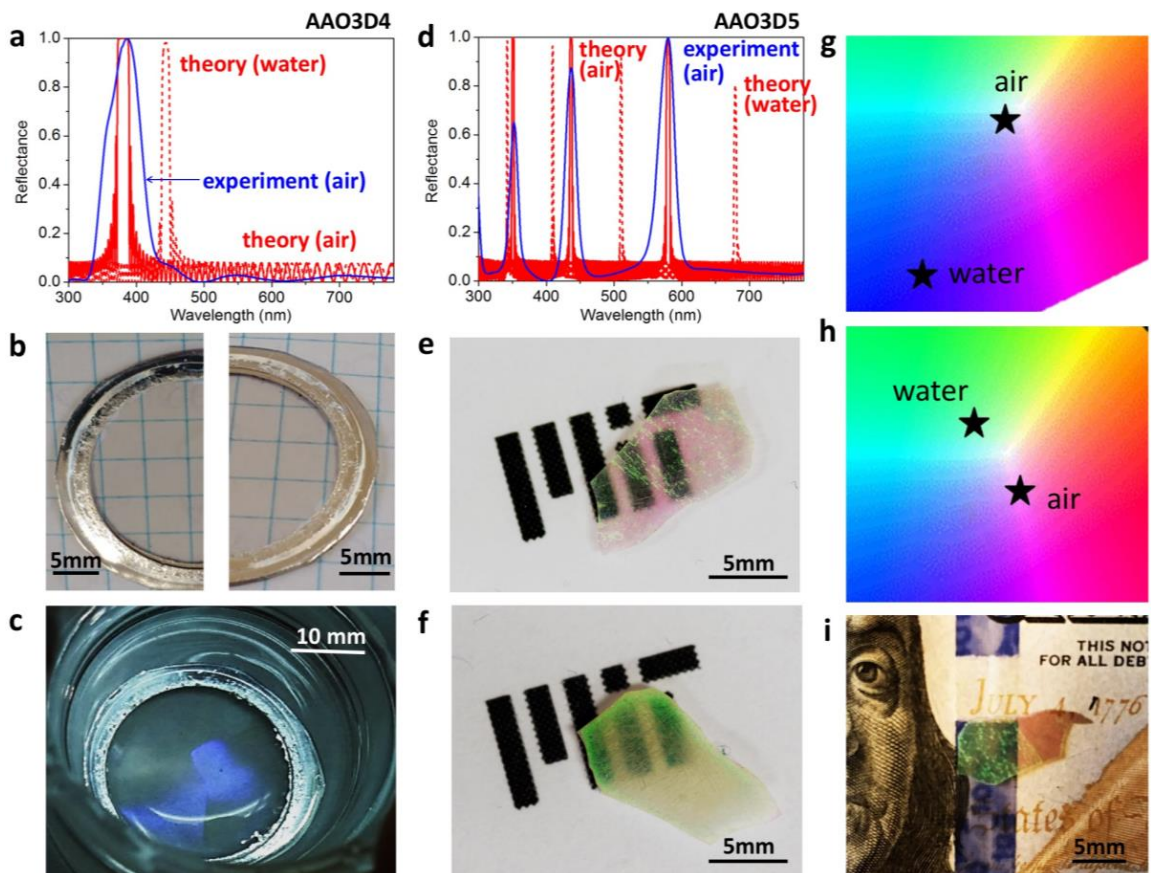

**Figure 4. Colorimetric sensing with porous 3D network metamaterials.** (a) The experimental (solid blue) and predicted (solid red) reflectance spectra of sample AAO3D4 at near-normal incidence. The dashed red line shows the predicted spectral shift upon sample immersion in water. (b,c) Photographs of sample AAO3D4 in air in (b) and in water in (c). (d-f) Same as (a-c) but for sample AAO3D5. (g,h) Colors of sample AAO3D4 (g) and AAO3D5 (h) in air and water on the CIE color diagram. (i) Color change of sample AAO3D5 from red to green on different substrates.

Summarizing, we have successfully engineered a wide range of structural colors in mesoporous alumina samples using internal periodicity, which illustrates the precision and robustness of the design methodology and the fabrication process. The fabricated samples are macroscopic in size, of around 25 mm in diameter for laboratory purposes, but the anodization method is solution-based, and can be carried out in much larger areas, making them useful for real-world applications. Neither low-throughput nano-patterning techniques nor precious metals are required to generate color, making these materials an attractive low-cost alternative to plasmonic colored nano-materials. Ultra-high porosity and strong spectrally- and environmentally-selective reflectance open up a vast field of possible applications for the mesoporous network meta-materials, including environmental monitoring, color filtering, food safety, homeland security, anti-counterfeiting, and healthcare.

## Methods

Samples were prepared from aluminum foils of 99.999% purity from Advent Research Materials, which were cleaned in four steps of 4 minutes of sonication with acetone, water, isopropanol and ethanol and then electropolished in ethanol and perchloric acid (3:1) for 4 minutes at 20 V. Then the aluminum foils are anodized in a 0.3 M sulfuric acid solution for 24 hours at 0°C with a voltage of 24 V. After this first anodization step, the formed alumina layer is removed by chemical etching (phosphoric acid 6% wt., chromic oxide 1.8% wt. and deionized water). The second anodization step consists of a pulsed anodization carried out in the same solution and at the same temperature as the first step, with periodic pulses at 25 V (which produces mild anodization), followed by shorter pulses at higher voltages (as those described in ref[49]). These higher voltages produce hard anodization layers. These layers are etched more quickly than the mild anodization layers, enabling the formation of the transversal pores and thus generation of the internal 3D structure of the samples. By changing the duration of the mild anodization step, we can produce different vertical distances between the horizontal layers with transverse pores. Then, the remaining aluminum substrate is etched ($CuCl_2$, HCl, and deionized water) and the barrier layer is dissolved at 30°C in a $H_3PO_4$ (10% wt.) solution. Finally, the samples are immersed in a similar, but less concentrated solution ($H_3PO_4$ 5% wt.), to reveal the transversal pores. A schematic of the fabrication method is shown in detail in supplementary Fig. S2. The total thickness of the samples ranged from about 30 to 100 microns, and spectral measurements together with SEM imaging revealed that more homogeneous etching of the samples was achieved if the thickness was about 30 microns. This is evident from the comparison of the experimental spectra of thicker samples AAO3D1 and AAO3D2 (thicknesses ~ 90 µm), which feature broadened spectra, and thinner samples AAO3D3-5 (30 µm).

High-resolution Scanning Electron Microscope (SEM) Images of the samples were taken with a Verios 460 from FEI®. Spectroscopy measurements (reflectivity as a function of the wavelength) were carried out with a Perkin Elmer – Lambda 950 Spectrophotometer UV/Vis using the Universal Reflectance Accessory (URA) module, between 200 and 800 nm in wavelength, with incidence angles of illumination and detection, with respect to the surface of the sample, of 30, 45 and 82 degrees (82° being the closest to the case of normal incidence that the equipment could measure). The reflectance measurements between 300nm-800nm were also carried out with the UV/Vis spectrophotometer of Agilent Cary 5000 using the integrating sphere. Photographs of the various mesoporous structures were taken with a Nikon D600 camera and a Nikkor 105mm f/2.8 AF-D microlens.

## Computational Methods

The predictions of the reflectance spectra of mesoporous metamaterial samples were done by using the semi-analytical transfer matrix method[59,60] (Supplementary Fig. S5). The reflection coefficient of a multi-layered structure $R = |r_1|^2$ can be calculated by a using simple recursive formula: $r_n = E_{n-}/E_{n+} = (\rho_n + r_{n+1}e^{-2ik_n d_n})/(1 + \rho_n r_{n+1}e^{-2ik_n d_n})$, $n = N, N-1, ..., 1$, where $\rho_n$ is the Fresnel reflection coefficient of the $n$-th material interface, $k_n$ in the normal component of the wavevector in the $n$-th medium, and $d_n$ is the thickness of the $n$-th layer. The recursion is initialized by setting $r_{N+1} = \rho_{N+1}$. The number of layers in all the calculated structures was equal to 150, i.e., they were composed of 75 periodically-repeating pairs of high-porosity and low-porosity layers. The spectral positions of the reflection bands (and thus the colorimetric response) do not depend on the number of layers in the structure (see supplementary Fig. S15), and $N$=150 was simply chosen to achieve high reflectance within each stopband.

Effective refractive index of porous dielectric was calculated by using the Maxwell-Garnett equation, $\varepsilon_{eff} = \varepsilon_m (2p(\varepsilon_a - \varepsilon_m) + \varepsilon_a + 2\varepsilon_m)/(2p(\varepsilon_m - \varepsilon_a) + \varepsilon_a + 2\varepsilon_m)$, where $\varepsilon_m$ is the permittivity of material (i.e., alumina), $\varepsilon_a$ is the permittivity of the ambient medium (e.g., air or water), and $p$ is the porosity of the material. The effect of the increased porosity of alumina on its reflective index is illustrated in Supplementary Fig. S4. The porosity values used in the modeling (65/80) were chosen by training the model on measured spectra of multiple fabricated samples to find the unique porosity values that match all the experimentally-observed spectral bands of all the samples in the training set. These values did not agree perfectly with the porosity estimates obtained from the SEM images of the samples (45/80 in our estimate). Possible explanation for this discrepancy may be the deviation of the bulk refractive index of our samples from the typical bulk value for alumina due to specific fabrication conditions. The porosity values of 65/80 were used in the model to design samples AAO3D1-AAO3D5, and the predicted spectra showed good agreement with the measured spectra of the fabricated structures (see supplementary Fig. S10).

To compute the color coordinates $(X, Y)$ from the reflectance spectra, we calculated the tristimulus values (x,y,z) by integrating the products of the reflectance coefficient $R(\lambda)$, the corresponding CIE color matching function (Fig. 2b), and the sunlight spectral power distribution $I(\lambda)$: $x = \int_{\lambda} R(\lambda)I(\lambda)\bar{X}(\lambda)d\lambda$, and subsequently normalized them as $X = x/(x + y + z)$ [53].

**Acknowledgments**: We acknowledge the service from the X-SEM Laboratory at IMM, funding from MINECO under project CSIC13-4E-1794 with support from EU (FEDER, FSE), and support from MIT S3TEC Center (an Energy Frontier Research Center funded by the Department of Energy, Office of Science, Basic Energy Sciences under Award # DE-FG02-09ER46577) for basic experimental infrastructure. M.S.M.G acknowledges financial support to the ERC POC 665634, the INFANTE project of CSIC 201550E072, and the Salvador Maradiaga fellowship from MINECO to work on sabbatical at Department of Mechanical Engineering, Massachusetts Institute of Technology.

**Author contributions:** A.R.C, O.C.C., M.M.G. addressed the fabrication of the 3D AAO mesoporous metamaterial network samples at IMN. A.R.C, Y.T., and G.N. performed optical characterization of the samples. A.R.C. performed also the SEM characterization. S.V.B. performed optical and colorimetric design of the meta-material. All the authors discussed and analyzed the results, O.C.C., M.M.G, and S.V.B. wrote the paper, G.C., S.V.B., and M.M.G. supervised the project. Corresponding authors emails: sborisk@mit.edu, marisol@imm.cnm.csic.es

**Supporting Information:** Refractive index and reflectance for normal incidence of bulk alumina, the CIE color space chromaticity diagram, pigment and structural color mixing diagrams, reflectance spectra of different samples, and a video showing the effect of wetting on the color change in a sample are included here.

# Supporting information

# Engineering a full gamut of structural colors in all-dielectric mesoporous network metamaterials

Alejandra Ruiz-Clavijo[1], Yoichiro Tsurimaki[2], Olga Caballero-Calero[1], George Ni[2], Gang Chen[2], Svetlana V. Boriskina[2]*, Marisol Martín-González[1,2]*

[1] IMN-Instituto de Micro y Nanotecnología (CNM-CSIC), Isaac Newton 8, PTM, E-28760 Tres Cantos, Madrid, Spain.
[2] Department of Mechanical Engineering, Massachusetts Institute of Technology, Cambridge, MA 02139, USA

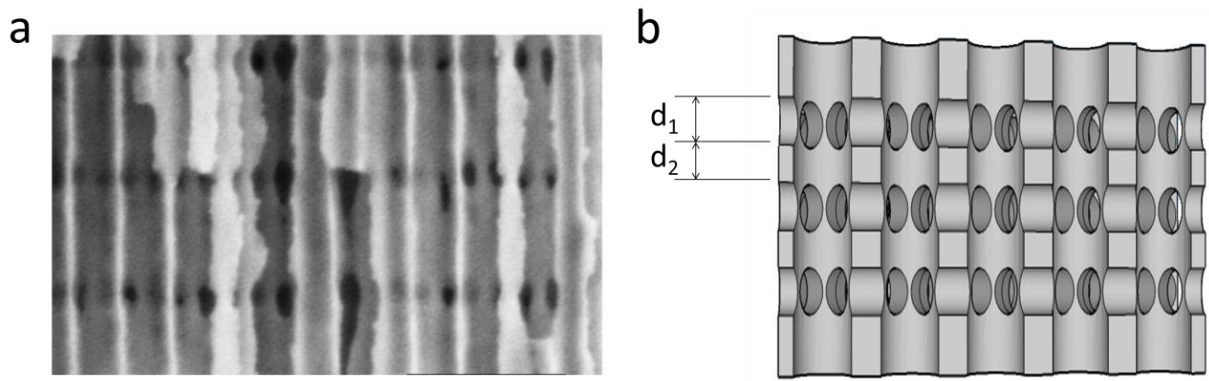

**Supplementary Fig. S1.** Cross view image (a) and schematic illustration (b) of a three-dimensional alumina membrane (3DAAO), where the high porosity layer with thickness $d_1$ and low porosity layer with thickness $d_1$ are defined.

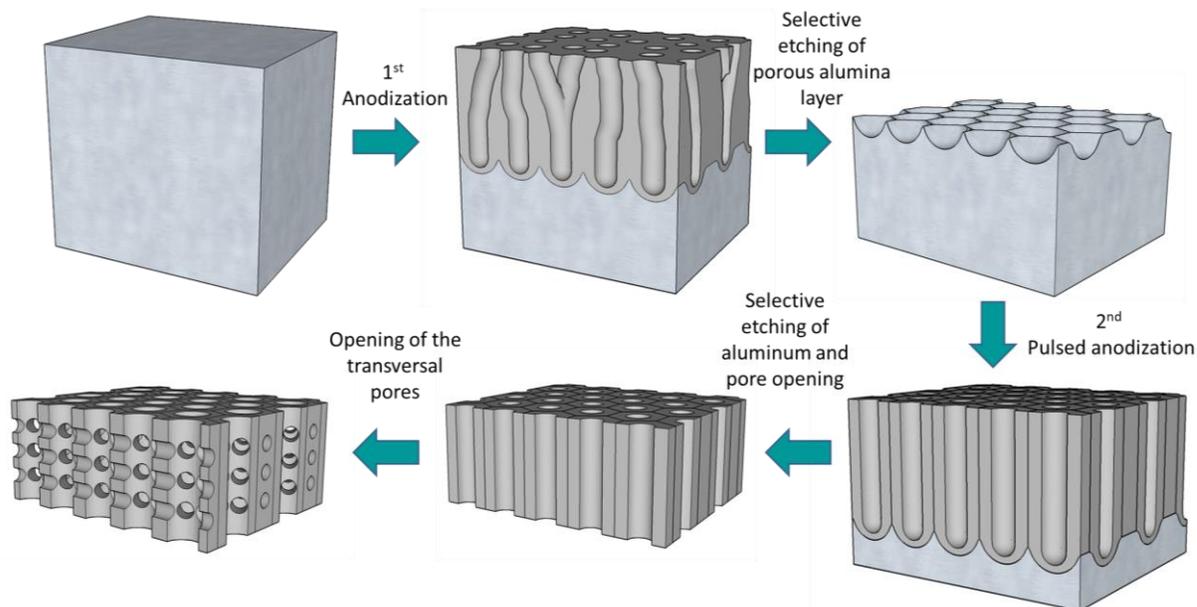

**Supplementary Fig. S2.** Schematic of the fabrication method used in the preparation of the 3D AAO porous metamaterial samples: Starting from pure aluminum (up-left), a first anodization is carried out, forming inhomogeneous pores at the surface, which get arranged in a hexagonal packaging after a certain anodization time (at a certain depth). Next, the alumina is removed by chemical etching, leaving an ordered patterned surface in the aluminum (up-right). Then, the second anodization is made, creating ordered pores. After the remaining aluminum and barrier layer removal (down-center image), the whole structure is immersed in phosphoric acid to reveal the transversal pores, created at the parts where the high voltages were applied in the second anodization step.

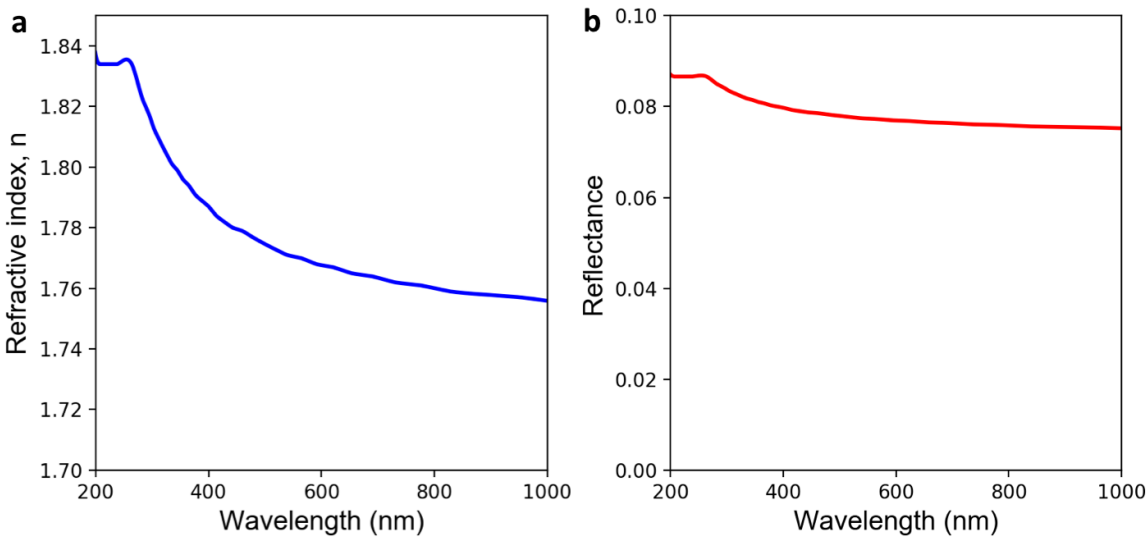

**Supplementary Fig. S3**. Refractive index (a) and reflectance (b) for normal incidence of bulk alumina ($Al_2O_3$) (*1*).

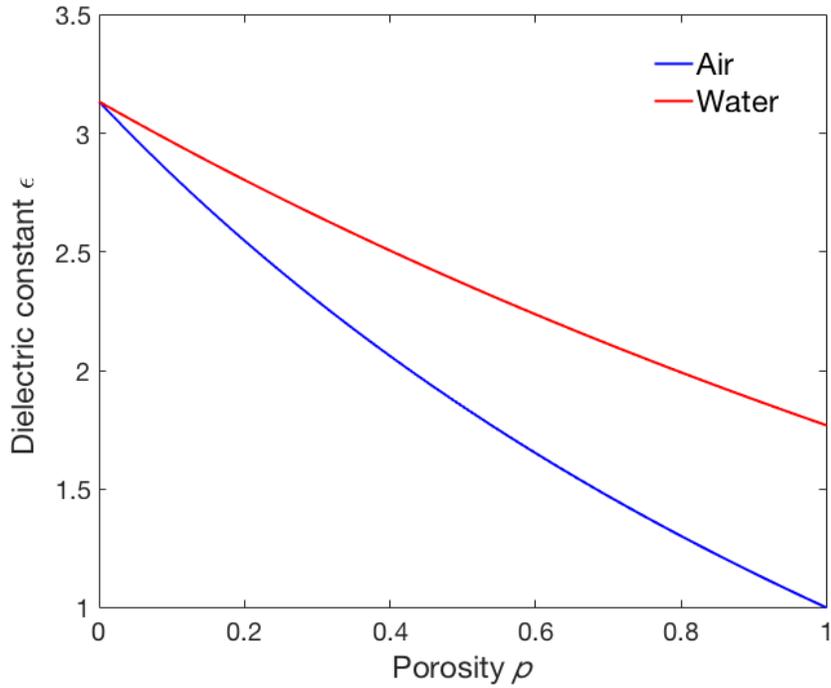

**Supplementary Fig. S4**. The effective dielectric constant of porous alumina calculated at a wavelength of 550nm via Maxwell-Garnett method as a function of material porosity. Two cases are considered: pores filled with air (blue line) and pores filled with water (red line). The dielectric constant of bulk $Al_2O_3$ at this frequency is 3.1329 ($n$=1.77, see also Fig. S3), and the dielectric constant of water is 1.777 ($n$=1.333).

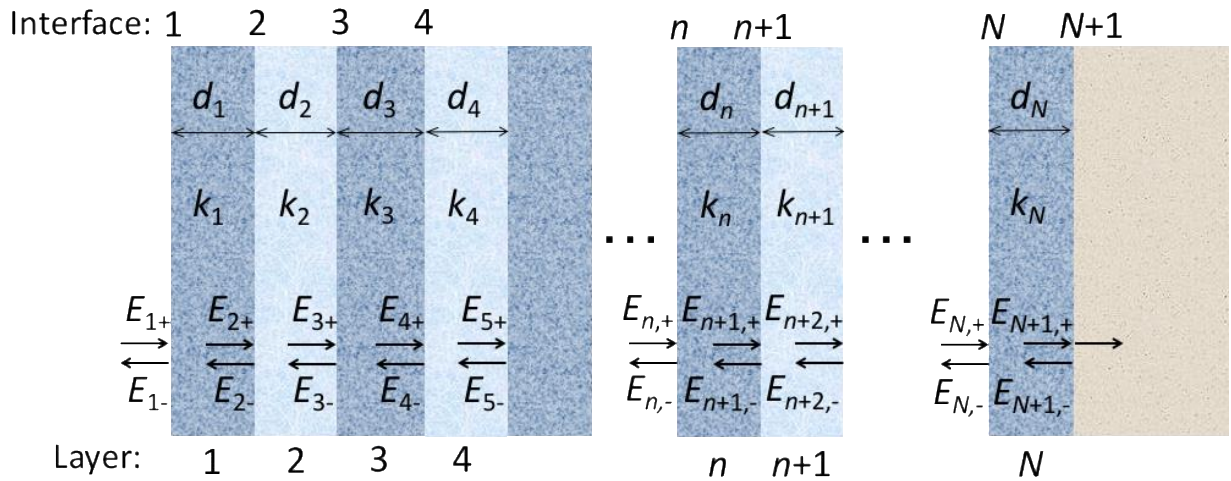

**Supplementary Fig. S5**. Schematic of the modeled multilayer structure with numbering convention of interfaces, layer parameters, as well as the strengths of the incident and reflected electric fields at each interface used in the Transfer Matrix Method calculations (see Methods).

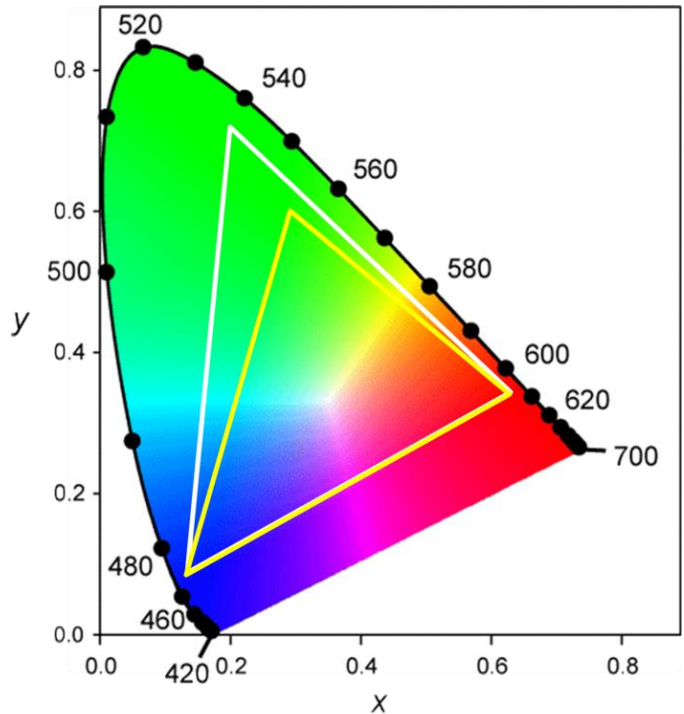

**Supplementary Fig. S6**. The CIE 1931 color space chromaticity diagram. The outer curved boundary is the spectral (or monochromatic) locus, with wavelengths shown in nanometers (image credit: Wikipedia). The CIE system characterizes colors by two-color coordinates X and Y, which correspond to a point on the chromaticity diagram. These coordinates are determined by the spectral power distribution of either emitted or reflected light and by the average sensitivity curves measured for the human eye (see Methods). The areas of the triangles illustrate the partial gamut of colors that can be matched by various combinations of red, green, and blue (RGB) in the color monitors. The yellow triangle covers the standard RGB (sRGB) gamut of colors represented in typical HDTV screens[2], and the white triangle covers the range of colors included in the Adobe RGB color space[3].

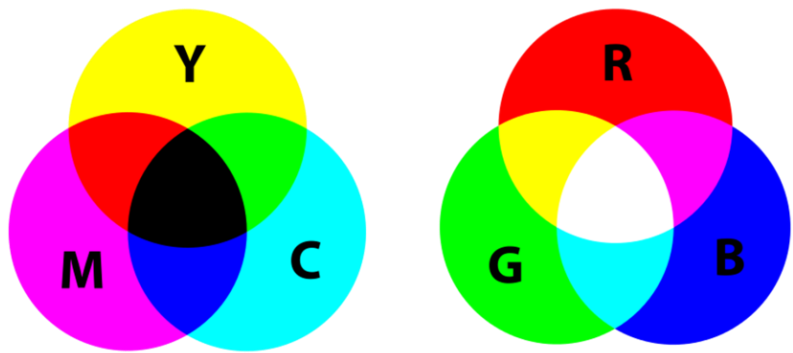

**Supplementary Fig. S7**. Pigment or additive (left) and structural or subtractive (right) color mixing diagrams. The sum of primary pigment colors yields black, while the addition of structural reflectance-based colors yields white (image credit: Wikipedia).

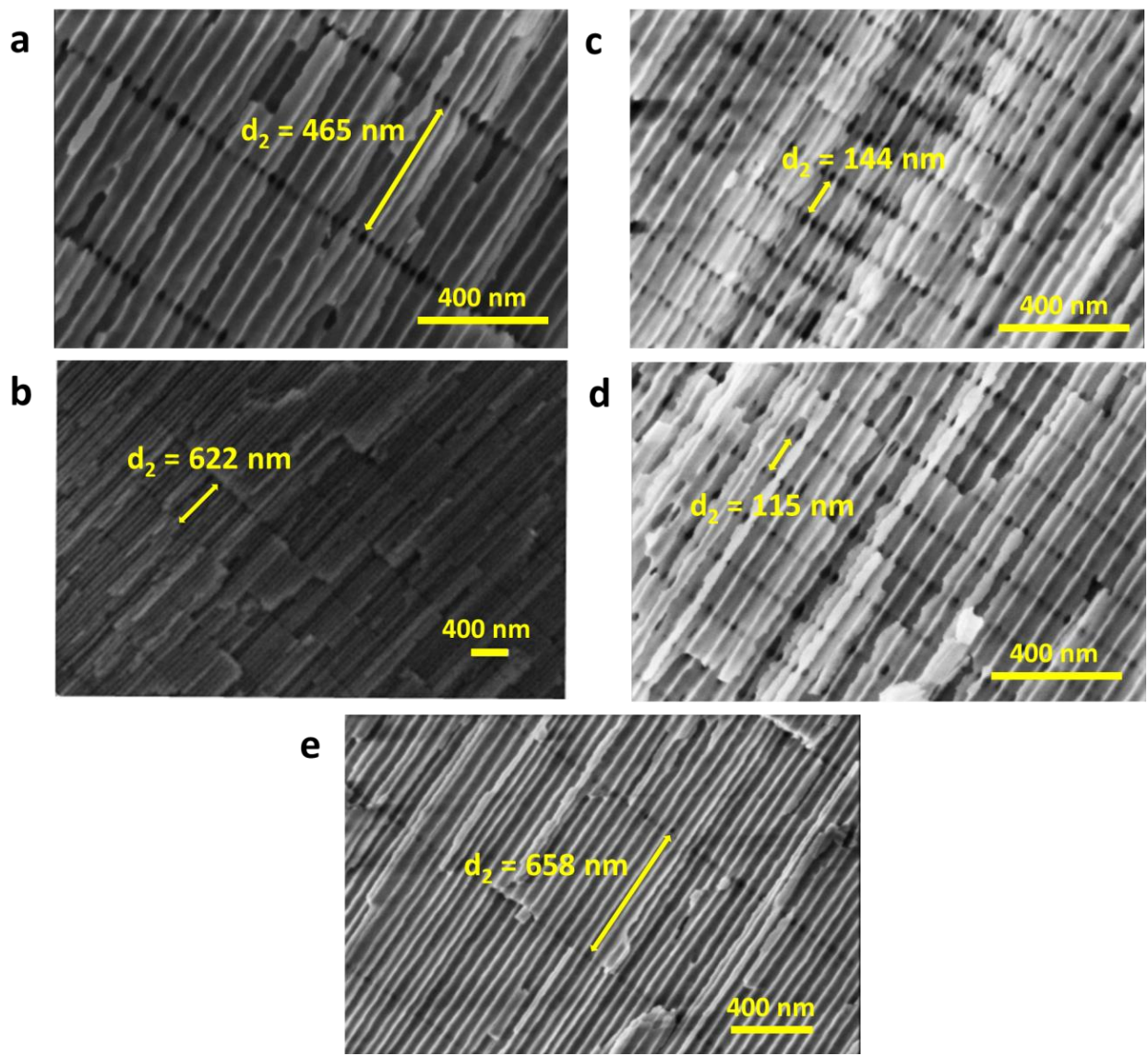

**Supplementary Fig. S8** Scanning electron microscope images of the structure of the fabricated samples shown in Table 2: a) AAO3D1, b) AAO3D2, c) AAO3D3, d) AAO3D4, and e) AAO3D5.

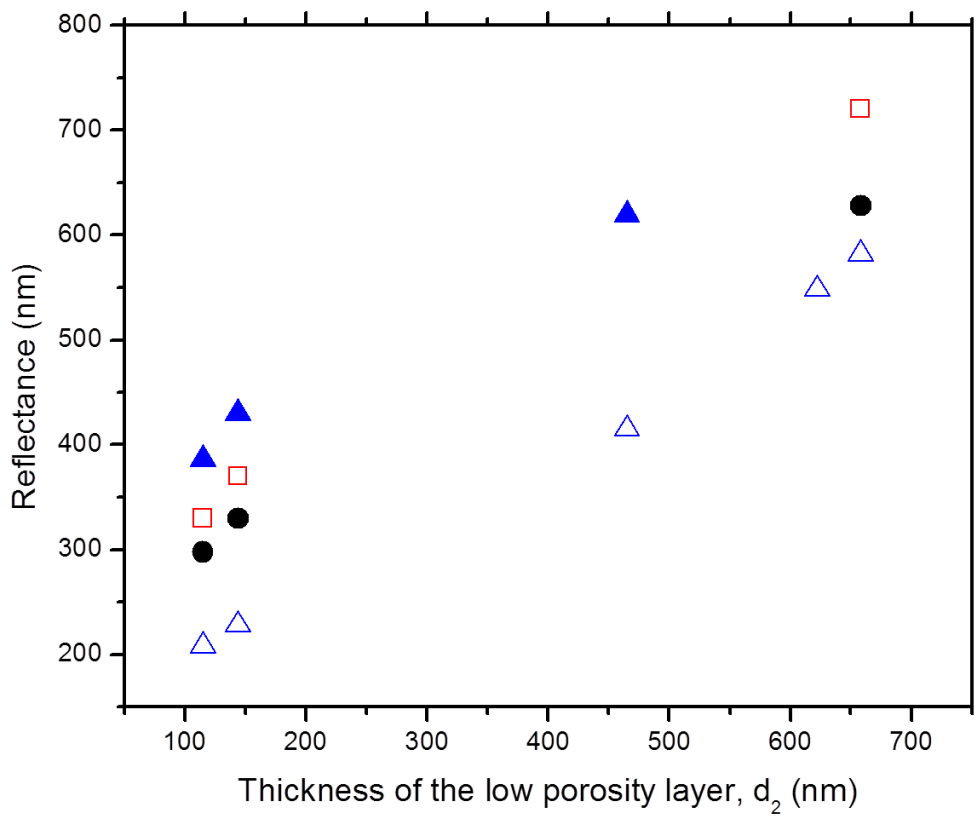

**Supplementary Fig. S9**. shows the wavelength at which the maximum reflectance of the longest-wavelength peak contributing to the structural color formation (i.e., the rightmost peak in the visible reflectivity spectra) appears in the samples of Table 2 as a function of the sample's periodicity for the different incidence angles measured, 30º (solid black circles), 45º (open red squares), and 82º (solid blue triangles). The open triangles represent the wavelength at which the next-order peak of the reflectivity spectra appears at 82º. Note that for the thickest low porosity layer, the low-order reflectance peaks lie in the infrared, and thus are not shown, given that they are out of the detector range.

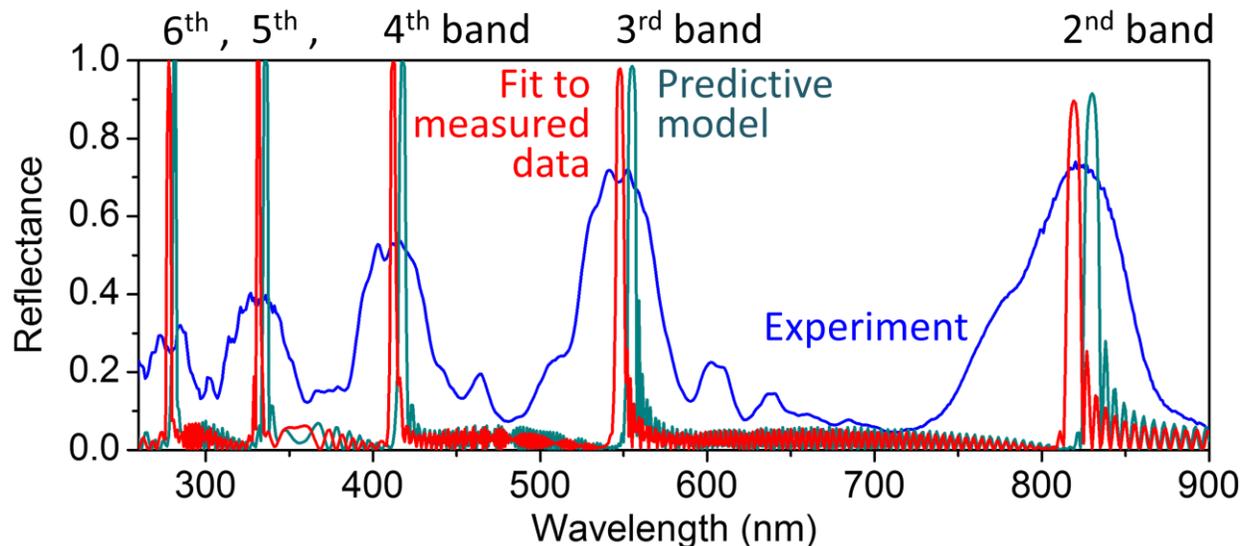

**Supplementary Fig. S10**. The reflectance spectrum of sample AAO3D2 across the ultra-violet, visible, and near-infrared parts of the electromagnetic spectrum. The teal line corresponds to the pre-fabrication model predictions for the sample with $d_1$=30nm, $d_2$=625nm, and porosities of 65% and 80%, respectively. The blue

line is the experimentally measured spectrum. The red line is the best numerical fit to the experimental spectra, calculated for the structure with $d_1$=32nm, $d_2$=622nm (measured), and porosities of 67% and 80% (fitted). Experimental higher-order peaks get progressively more deformed and reduced in the magnitude due to fabrication imperfections.

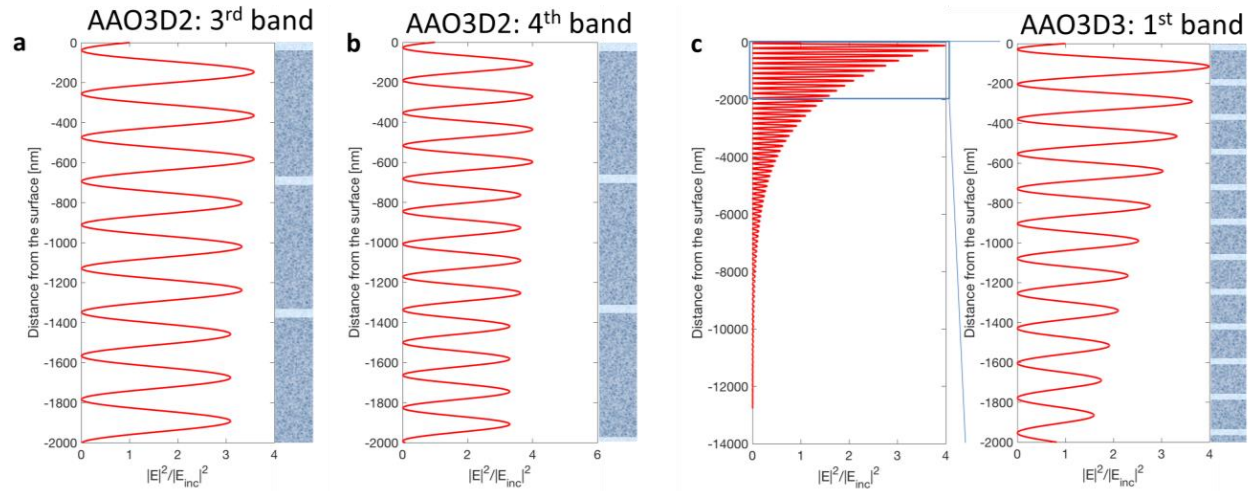

**Supplementary Fig. S11**. (a,b) The electric field distributions in the sample AAO3D2 at wavelengths 555 nm (a) and 418 nm (b), corresponding to the third and fourth stop bands of the porous metamaterial stack. The geometry of the multi-layered metamaterial stack is schematically shown to the right of each plot. The internal field variations within each layer are on the scale smaller than the pore sizes. This results in the spectral broadening of these bands and their deviation from the theoretical predictions made under the assumption of uniform layers with effective permittivity values. (c) The electric field distribution in the sample AAO3D3 at wavelength 441 nm, corresponding to the first-order stop bands of this sample is shown for comparison. Comparison of panels (a,b) and (c) reveals deeper penetration lengths of the electric field into the sample at frequencies corresponding to the higher-order bands. This, in combination with multiple field variations within each low-porosity layer, makes high-order bands more sensitive to the fabrication details, including pore presence, pore non-uniformity, variations in the layer thickness, etc.

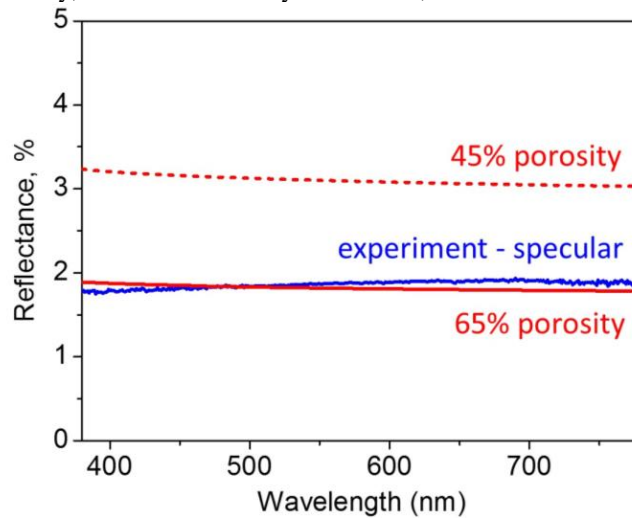

**Supplementary Fig. S12**. Reflectance spectra of a AAO sample without transverse pores. The incident angle is 8° from the normal. Blue: experimentally measured specular reflection spectrum. Solid red: calculated spectrum of a sample with 65% uniform porosity. Dashed red: calculated spectrum with 45% porosity value.

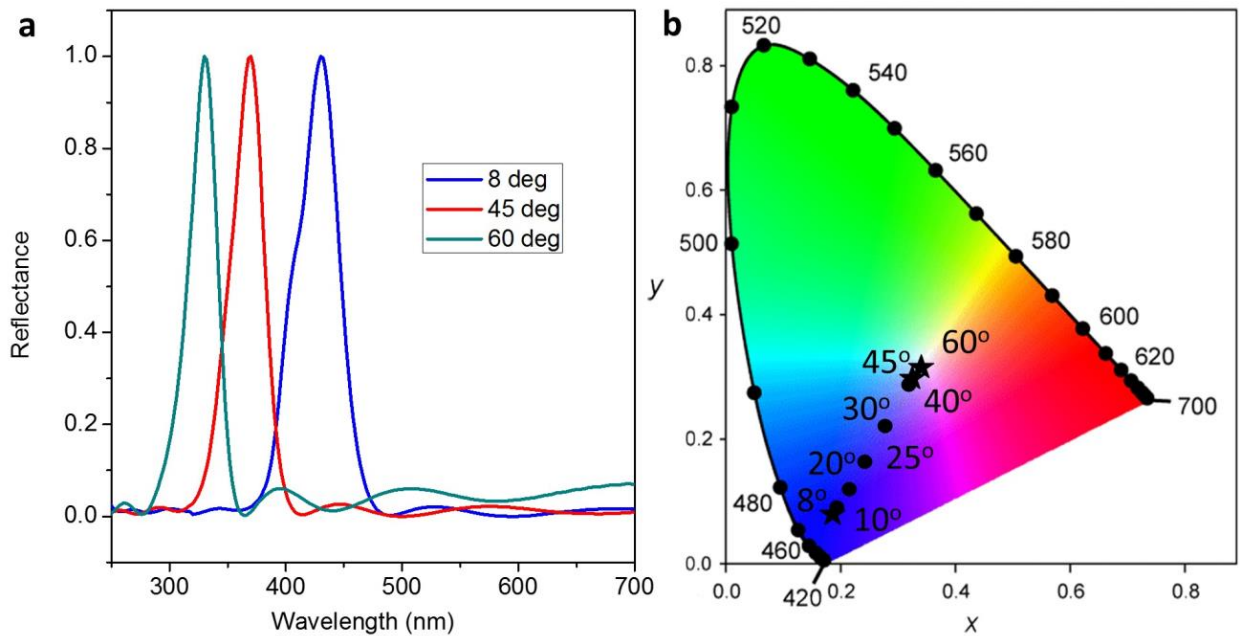

**Supplementary Fig. S13**. (a) Experimental reflectance spectra of sample AAO3D3 measured at 8° (blue), 45° (red), and 60° (teal) to normal. (b) The points on the CIE chromaticity diagram corresponding to the experimental (stars) and calculated (dots) reflectance spectra at different angles measured from normal. The points move towards the center of the diagram with the increased angle of observation, resulting in the colorimetric response of the sample weakening and eventually dissapering at an angle of about 40°.

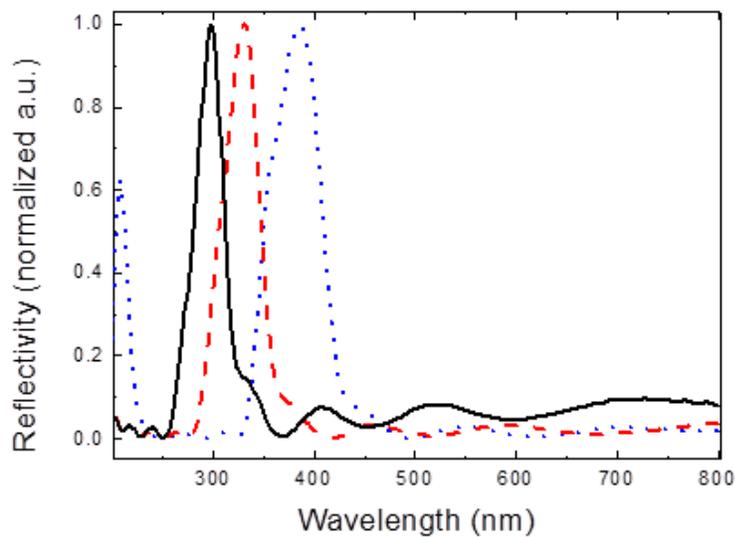

**Supplementary Fig. S14**. Experimental reflectance spectra of sample AAO3D4 measured at 8° (dotted blue), 45° (dashed red), and 60° (solid black) to normal.

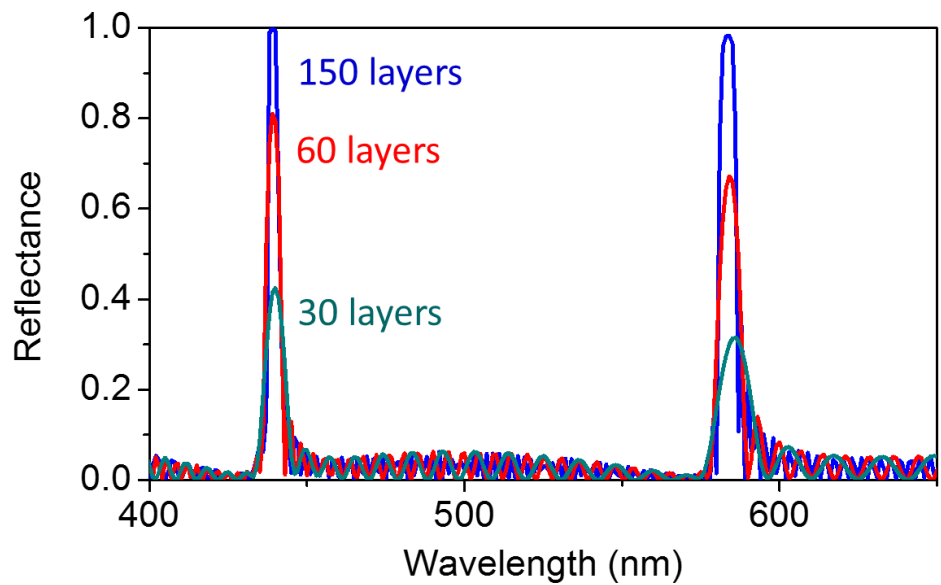

**Supplementary Fig. S15**. Calculated reflectance spectra for sample AAO3D5 obtained by using an increasing number of layers *N* (as labeled on the plot). Typical for the Bragg interference mechanism in periodic solids, the reflectance bands become more pronounced with the increase of the structure thickness, while their spectral positions remain virtually unchanged.

**Supplementary video V1** shows the effect of wetting on the color change in sample A5.